\documentclass[twocolumn,prl,showpacs]{revtex4}
\usepackage{dcolumn,times,epsfig,color}

\newcommand{\Label}[1]{\label{#1}}
\newcommand{\figref}[1]{Fig.\,\ref{#1}}
\newcommand{\GPE}{Gross-Pitaevskii equation}
\newcommand{\BEC}{Bose-Einstein condensate}

\begin{document}
\title{Quantum turbulence in condensate collisions: an application of 
the classical field method}
\author{A. A. Norrie$^1$, R. J. Ballagh$^1$ and C. W. Gardiner$^2$}
\affiliation{
$^1$Physics Department, University of Otago, Dunedin, New Zealand \\
$^2$School of Chemical and Physical Sciences, Victoria University of Wellington,
New Zealand}
\begin{abstract}
We apply the classical field method to simulate the production of correlated
atoms during the collision of two Bose-Einstein condensates. Our
non-perturbative method includes the effect of quantum noise, and provides for
the first time a theoretical description of collisions of high density
condensates with very large out-scattered fractions. Quantum correlation
functions for the scattered atoms are calculated from a single simulation, and
show that the correlation between pairs of atoms of opposite momentum is rather
small. We also predict the existence of \emph{quantum turbulence} in the field
of the scattered atoms---a property which should be straightforwardly
measurable.
\end{abstract}
\pacs{03.75.Kk, 05.10.Gg, 34.50.-s \vspace{-0.5cm}}
\maketitle


In the same way as a classical electromagnetic field obeying Maxwell's
equations arises as an assembly of photons all in the same quantum state, a
\BEC, composed of Bosonic atoms all in the same quantum state, behaves very
much like a classical field $\Psi({\bf x},t)$, whose equation of motion is the
\GPE
\begin{equation}
\Label{GPE in coordinate space}
i\hbar \frac{\partial \Psi \left( \mathbf{x},t\right) }{\partial t}=
\left[ -\frac{\hbar ^{2}\nabla ^{2}}{2m}+\frac{4\pi \hbar ^{2}a}{m}
\left| \Psi \left( \mathbf{x},t\right) \right| ^{2}\right] 
\Psi \left( \mathbf{x},t\right).
\end{equation}
Nevertheless, there are phenomena in which the quantized nature of this field
is important---for example, when two Bose-Einstein condensates collide at a
sufficiently high velocity, a \emph{halo} of elastically scattered atoms is
produced \cite{Chikkatur2000a,Katz2002a,Vogels2002a}. The \GPE\ with initial
conditions corresponding to two \BEC s does not predict this scattering---it is
a direct effect of the fact that the quantized field consists of interacting
particles.  


Theoretical descriptions of this phenomenon fall into two groups. In the 
first, the \GPE s for the two condensate wavepackets are modified (either
phenomenologically \cite{Band2000a}, or on the basis of a method of
approximating quantum field theory \cite{Koehler2002a}) to give an elastic
scattering loss term. While these methods yield equations of motion which allow
for depletion of the condensate wavefunctions, they do not include a
description of the scattered atoms, and hence cannot describe the effects of
bosonically stimulated loss. In the second class of treatments 
\cite{Bach2002a,Yurovsky2002a} the quantum field theory is linearized about the
condensate, yielding equations of motion linear in the fluctuation operators.
This method shows that the process is essentially one of four-wave mixing
between the two condensate fields and pairs of quantized
fluctuations---however, as a linearized theory, it can deal only with
perturbatively small amounts of scattering and cannot simultaneously account
for depletion of the condensate. Both of these formalisms are valid only in the
limit of weak scattering, but fail for large scattered fractions, such as we
treat in this paper.


In this work we will show that a treatment in terms of a \emph{classical field}
with added \emph{quantum fluctuations} is not only able to produce the
scattering halo, but also predicts a hitherto unobserved phenomenon, which we
shall call \emph{quantum turbulence}, in the resulting halo, which should be
straightforwardly observable. This method is able to describe: a) the evolution
(including large depletion) of the condensate wavefunction as the scattering
progresses; b) the full quantum correlations originating from the nonlinear
amplification of the quantum vacuum fluctuations.


The \emph{classical field method} can be formulated as follows:

\noindent i) The system is described by a $c$-number field amplitude
$\Psi(\mathbf{x},t)$ which satisfies the Gross-Pitaevskii equation (\ref{GPE
in coordinate space}) and has a mode expansion
\begin{equation}
\Psi \left( \mathbf{x},t\right) = \frac{1}{\sqrt{V}}\sum ^{M}_{j=1}
\alpha _{j}\left( t\right) e^{i\mathbf{k}_{j}\cdot \mathbf{x}},
\Label{Psi from Phi}
\end{equation}
where $\alpha_{j}\left( t \right)$ is the amplitude for the mode with
wavevector $\mathbf{k}_{j}$ and $V=L_{x}L_{y}L_{z}$ is the volume contained
within the periodic boundaries in coordinate space.

\noindent ii) The quantum fluctuations are introduced in the \emph{initial
condition} for $\Psi(\mathbf{x},t)$, which is given by the sum $ \Psi \left(
\mathbf{x},t=0\right) =\psi \left( \mathbf{x}\right) + \chi \left(
\mathbf{x}\right) $, where $\psi\left(  \mathbf{x}\right)$  and $\chi\left( 
\mathbf{x}\right)$ are respectively the real and virtual particle fields. We
express the field of virtual particles using $\chi \left( {\bf x} \right) =
\sum_{j=1}^M \chi_j \exp \left( i {\bf k}_j \cdot {\bf x} \right) / \sqrt{V} $,
where the amplitude in each mode is Gaussian with the properties
\begin{eqnarray}
\Label{Noise properties}
\left\langle \chi _{i}^{*}\chi _{j}\right\rangle =
\frac{1}{2}\delta _{ij},\qquad
\left\langle \chi _{i}\chi _{j}\right\rangle =  0.
\end{eqnarray}
The mean value of the total virtual population is thus $M/2$.


The fundamental basis for the classical field method is the representation of
the system of many bosons by means of a Wigner function description, which for
large occupations is equivalent to this form
\cite{Steel1998a,Stoof1999a,Duine2001a,Stoof2001a,Davis2001a,Davis2001b,
Sinatra2001a,Goral2001a,Gardiner2002a,Davis2002a,Goral2002a,Schmidt2003a,
Polkovnikov2003a} (the truncated Wigner approach). Heuristically, the
introduction of virtual particles via (\ref{Noise properties}) can be viewed as
adding half of one particle per mode, corresponding to the zero-point
occupation of the ground state of the harmonic oscillator which represents each
mode. The choice of initial condition (\ref{Noise properties}) is correct where
$\psi({\bf x})=0$, but generates a slightly heated and nonequilibrium
condensate where $\psi({\bf x})\ne 0$ \cite{Steel1998a,Sinatra2000a}.  This
will make very little difference to the results of the calculations, which
require that the \emph{vacuum} be represented accurately---furthermore, in
practice the difference between the state thus represented and a pure
condensate is very much less than the experimental uncertainty.


The pseudopotential approximation used in Eq.(\ref{GPE in coordinate space})
results from the elimination of modes with momenta larger than a certain
cutoff. In the work of \cite{Band2000a,Koehler2002a} this cutoff effectively
leaves \emph{two} sets of modes, centered around the mean momenta of each of
the individual colliding wavepackets, and mutual scattering appears explicitly.
In contrast, we use a single spherical momentum space cutoff which is large
enough to include all relevant momenta for the scattering process, but small
enough for the pseudopotential approximation to retain its validity. This is
possible since the essential requirements for the elimination of the higher
modes are that these modes should have no occupation, and that the wavelength
at which the cutoff is made should be larger than the range of the interatomic
potential. The field is evolved using the \GPE\ within a projection formalism
similar to that introduced in \cite{Davis2001b}.


For this problem, we write the \GPE\ for each mode as
\begin{equation}
\Label{GPE for modes} 
i\hbar \frac{\partial \alpha _{j} }{\partial t} =  
\frac{\hbar ^{2}k_{j}^{2}}{2m}\alpha _{j}
+ \frac{4\pi \hbar ^{2}a}{mV}\sum_{rst}\alpha_{r}^{*} \alpha_{s} \alpha_{t}
\delta_{jr,st},
\end{equation}
where the function $\delta_{jr,st}$ is unity for $\mathbf{k}_{j}+\mathbf{k}
_{r}-\mathbf{k}_{s}-\mathbf{k}_{t}=0$ and zero otherwise. The projection is
implemented by the requirement that all of $|\mathbf{k}_i| ,|\mathbf{k}_r|
,|\mathbf{k}_s| ,|\mathbf{k}_t| $ be less than $K_{\rm max} $, the wavenumber
which sets the momentum space cutoff.

The nonlinearity present in the Gross-Pitaevskii equation gives rise to
pairwise interactions of the type $j+r \leftrightarrow s+t$, but only if the
amplitudes $\alpha_{r}$, $\alpha_{s}$ and $\alpha_{t}$ are all non-zero. As
noted in \cite{Vogels2002a}, this mechanism can be viewed as a non-degenerate
parametric oscillator, as found in nonlinear and quantum optics
\cite{Walls1995a}. In the classical field formulation, the noise terms $\chi_j$
ensure that \emph{all} modes do have non-zero occupation. Vacuum modes, in
which the occupation is initially virtual, can therefore develop a macroscopic
population as a result of the interaction. Thus the scattering in our formalism
arises naturally as a result of the classical field method.


The usual derivation of the classical field formulation via the truncated 
Wigner function
\cite{Gardiner1999a,Sinatra2001a,Gardiner2002a,Polkovnikov2003a} requires that
all relevant occupations be large, and does not immediately seem to be
satisfied here since we are considering modes whose only occupation is the
virtual population. However, for this situation, in which the most significant
contributions come from terms in which two of the $\alpha_{j}$ amplitudes in 
(\ref{GPE for modes}) are macroscopically occupied, it is still possible to
prove the negligibility of all terms with third-order derivatives in the
Fokker-Planck equation which represents the \emph{exact} equation of motion in
the Wigner representation, and this is the condition for the validity of the
classical field method. This result will be published elsewhere.


\begin{figure}
\begin{center}
\includegraphics{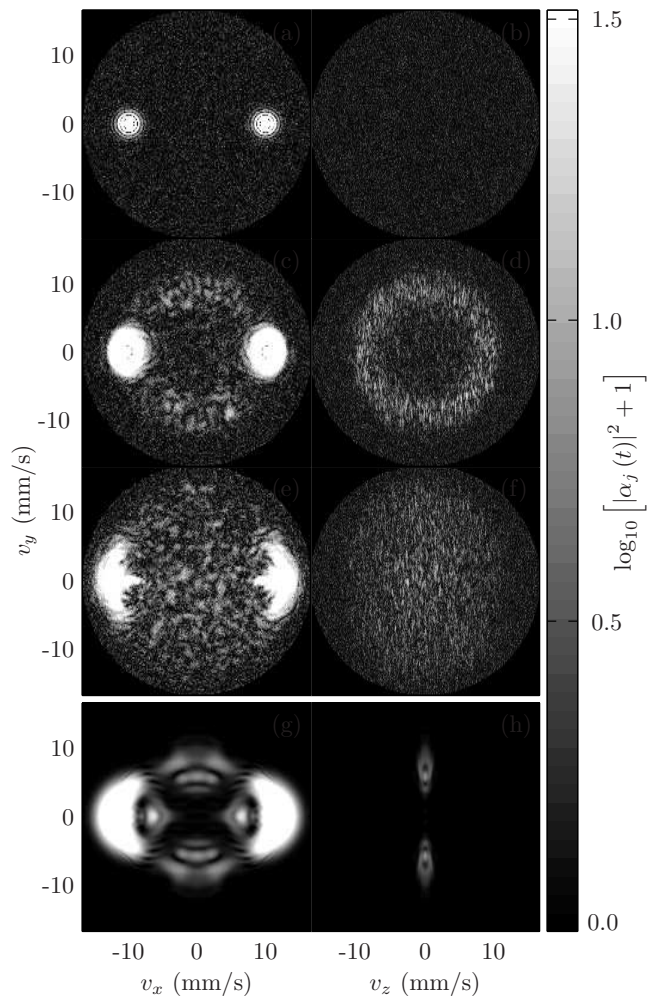}
\end{center}
\vspace{-0.5cm}
\caption{\label{Figure - slices} (a--f): Velocity mode populations on the
planes $v_z = 0 $ (left) and $v_x = 0$ (right) for the collision of packets
generated from a six million atom condensate at $t=0$ (top), $t = 0.5$ ms
(middle) and $t = 2.0$ ms (bottom). The spherical momentum cutoff is clearly
visible in the upper plots due to the presence of quantum fluctuations. 
(g--h): Mode populations at $t = 2.0$ ms for an identical collision excluding
vacuum noise.}
\vspace{-0.5cm}
\end{figure}


We consider collisions between two equally populated wavepackets derived from a
single trapped condensate using a Bragg grating of very short duration compared
to the overlap time of the packets \cite{Kozuma1999a,Ovchinnikov1999a}. Thus an
appropriate initial state, $\psi\left( \mathbf{x}\right)$ of the real particles
is the sum of two distinct momentum wavepackets with the same spatial envelope:
\begin{equation}
\Label{Initial state}
\psi \left( \mathbf{x}\right) = \frac{1}{\sqrt{2}} \psi_{0}
\left( \mathbf{x}\right) 
\left[ e^{i\mathbf{k}_{+}\cdot \mathbf{x}} + e^{i\mathbf{k}_{-}\cdot 
\mathbf{x}}
\right],
\end{equation}
where $\psi_{0}\left( \mathbf{x}\right) $ is the stationary solution of the
Gross-Pitaevskii equation for the $N_0$ atom condensate in a trap and
$\mathbf{k}_{\pm} = \left( \pm mv_c/\hbar,0,0 \right)$ are the wavepackets'
normalized amplitudes and wavevectors. For this paper we choose the case of a
$^{23}\mathrm{Na}$ condensate in an axially symmetric trap with
$\omega_{r}=2\pi\times80$ Hz and $\omega_{z}=2\pi\times20$ Hz, as used in
\cite{Vogels2002a}, but with a maximum of $6\times10^{6}$ atoms, compared to
the maximum number $3\times10^{7}$ used in the experiment. Immediately
following the application of the Bragg pulses ($t=0$) the confining potential
is removed, so that the evolution takes place in free space.


The velocity space population distribution for the collision of packets from an
initial condensate of $6\times 10^6$ atoms with $v_c = 10$ mm/s is shown in
\figref{Figure - slices}(a-f) for a sequence of times. This figure shows:

\noindent
i) The original wavepackets broadening and changing shape from circular in
projection to crescent shaped.  

\noindent
ii) The generation of a circular feature---\emph{the scattering
halo}---centered at the system center of mass velocity with a radius in
velocity space approximately equal to (but larger than) $v_{c}$.

The scattering halo is characterized by patches of high mode population, the
\emph{phase grains}, whose size and aspect ratio are very similar to those of
the original condensate packets (in velocity space). This characteristic size
can be understood in terms of the parametric amplification of vacuum
fluctuations described by Eq.(\ref{GPE for modes}). Pairs of vacuum modes $j$
and $r$ at each end of a diameter of the energy conservation sphere are
selected preferentially for growth (or loss) if their phases are appropriate.
The sum in Eq.(\ref{GPE for modes}) convolves the noise field near mode $r$
with the initial condensate packets, and if mode $r$ grows preferentially, it
drives the growth of all modes near $j$ within a volume of the size of the
condensate velocity wavefunction.

In the weakly populated regions separating the phase grains, analysis of our
data reveals a large number of vortices, which is indicative of a turbulent
velocity field and arises directly from the parametric amplification of the
vacuum fluctuations. This justifies our identification of the behaviour as
\emph{quantum turbulence}. (An analogous amplification of
\emph{electromagnetic} vacuum fluctuations was achieved in 1982 using a
Josephson junction \cite{Koch1982a}.)

In contrast, a simulation performed without the vacuum fluctuations generates a
simple modulational instability, as shown in \figref{Figure - slices}(g--h). 
The fluctuations are equivalent to $3\times10^6$ additional particles, and thus
their inclusion represents a significantly altered initial condition from that
given by omitting them altogether.

\begin{figure}
\begin{center}
\includegraphics{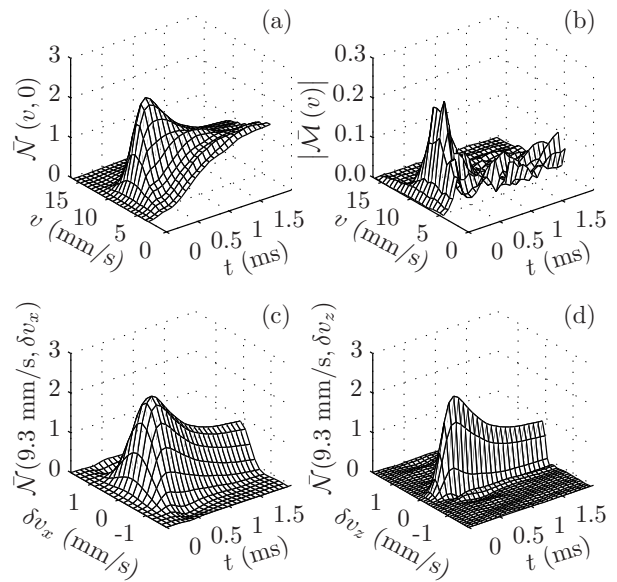}
\end{center}
\vspace{-0.5cm}
\caption{\label{Figure - correlation functions} Correlation functions for the
six million atom simulation averaged over the data collection region as
described in the text: (a) average mode population; (b) pair creation
correlation; (c,d) amplitude autocorrelation.}
\vspace{-0.5cm}
\end{figure}

As the halo generation process is analogous to parametric amplification, it is
expected that modes of opposite velocity in the halo will display correlations,
and that modes of similar velocity will similarly display correlations due to
the finite size of the phase grains. Calculating these correlation functions
should properly be performed on an ensemble of simulations. However, for this
situation which, away from the condensate packets, displays a high degree of
spherical homogeneity in velocity space, we can use a single simulation and
average over modes in velocity space which have similar speeds and which are
not occupied by the condensate packets. To this end we define the \emph{data
collection region}, consisting of those modes whose polar angles, $\varphi_j$,
lie between $\pi/4$ and $3\pi/4$ to the positive $v_x$ axis. Clearly such a
region would also be needed experimentally. Within this region then, we define
the correlation functions
\begin{eqnarray}
\!\!\!\! \bar{\mathcal{N}} \left( v,0 \right) & = & 
\left \langle \left| \alpha_j \right|^2 \right \rangle_{\left| {\bf v} \right|
= v} - \frac{1}{2}, \label{average mode population} \\
\!\!\!\! \bar{\mathcal{N}} \left( v,\delta {\bf v} \right) & = & 
{\rm Re} \left[ \left \langle \alpha \left( {\bf v} \right)^* \alpha
\left( {\bf v} + \delta {\bf v} \right) \right \rangle
_{\left| {\bf v} \right| = v} \right]  - \frac{\delta_{\delta {\bf v},0}}{2} ,
\label{displaced population correlation function} \\
\!\!\!\! \bar{\mathcal{M}} \left( v \right) & = &
\left \langle \alpha_i \alpha_j \right \rangle_{\left| {\bf v} \right| = v},
\label{pair creation correlation function}
\end{eqnarray}
as the average non-condensate mode population (excluding virtual particles),
the amplitude autocorrelation function and the pair creation correlation
function respectively. The correlation functions for the six million atom
simulation are shown through time in \figref{Figure - correlation functions}.


From \figref{Figure - correlation functions}(a), we observe that population
builds up initially near $v=10$ mm/s, peaking at $t=0.52$ ms for $v=9.3$ mm/s.
The subsequent redistribution of population within the halo modes occurs via
inter-halo scattering events and can be considered as a thermalisation process.

\figref{Figure - correlation functions}(b) shows the pair creation correlation
for the paired modes, $\mathbf{v}_{i}=-\mathbf{v}_{j}$ and shows that a
definite, although small ($\approx 10$\%), correlation develops, peaking at
$t=0.30$ ms and then decaying as further scattering events occur.

\figref{Figure - correlation functions}(c,d) show the velocity space amplitude
autocorrelation functions at $v=9.3$ mm/s. Fitting a Gaussian in $\delta
v_{x,y,z}$ to obtain the FWHM correlation lengths, we find that from the time
that population is established (at about $t=0.05$ ms) the ratio of the $\delta
v_{x}$ and $\delta v_{y}$ correlation lengths to the $\delta v_{z}$ length
remain close to 4. The individual lengths initially grow and peak at $t=0.25$
ms (where they are 1.2 mm/s in the $\delta v_x$ and $\delta v_y$ directions,
and 0.3 mm/s in the $\delta v_{z}$ direction) and then decay slowly. These can
be compared to the FWHM of the original velocity space condensate wavefunction,
which is 0.7 mm/s in the $v_{x}$ and $v_{y}$ directions and 0.18 mm/s in the
$v_z$ direction. This correlation behavior is a quantitative measure of the
size of the phase grains, and we estimate the average number of atoms in a
coherent patch to be 150 at $t = 0.25$ ms for $v = 9.3$ mm/s.

\begin{figure}
\begin{center}
\includegraphics{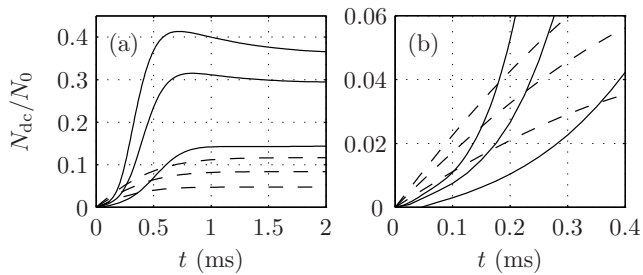}
\end{center}
\vspace{-0.5cm}
\caption{\label{Figure - populations} Normalised data collection region
populations for the classical field simulations (sold lines) and complex
scattering length simulations (dashed lines). Initial condensate numbers are
(lowest to highest) $\left( 1,3,6 \right) \times 10^6$, all with $v_c = 10$
mm/s. Subplot (b) shows the early times of (a).}
\vspace{-0.6cm}
\end{figure}

In \figref{Figure - populations} we examine the time evolution of the real
particle population in the data collection region, defined using
\begin{equation}
\label{DC region population}
N_{\rm dc} = \sum_j \left( \left| \alpha_j \right|^2 - \frac{1}{2} \right),
\hspace{0.5 cm} \frac{\pi}{4} \leq \varphi_j \leq \frac{3\pi}{4},
\end{equation}
for varying initial condensate number. For larger condensate numbers a larger
fraction is scattered, and the scattering is completed earlier. The decrease in
population at later times is due to interhalo scattering events redistributing
population outside the data collection region.

For comparison, we plot the fraction of condensate population lost into the data
collection region (assuming isotropic loss) during an identical collision using
the complex scattering length method of \cite{Band2000a}. The most significant
difference using our method is a considerably larger out-scattered fraction,
because Bosonic stimulation is included. Additionally, the scattering model of
\cite{Band2000a} is essentially based on Fermi's ``golden rule'', which results
in exact kinetic energy conservation at all times, and gives an initial
\emph{linear} growth of the out-scattered fraction. However, our method shows
an initial \emph{quadratic} growth rate, because the timescale over which one
must average to derive the kinetic energy conservation implicit in the golden
rule (say to an accuracy of about 10\%) in this case is about $0.5{\rm s}$. As
a result of both of these effects, the rate of population loss, instead of being
greatest at $t=0$, where the wavepackets are maximally overlapped, is greatest
at about $t=0.3$ -- $0.5{\rm ms}$.


The major loss mechanism in a BEC is that of three body recombination
\cite{Fedichev1996a,Ketterle1998a}, but this can have very little effect on the
results. Using the experimentally obtained loss rate, we estimate that in the
time taken ($\approx 2{\rm ms}$) for an experiment, no more than 5 atoms would
be lost from the system. Implementing a phenomenological three-body loss into
our simulations shows no change in the dynamics until we increase the loss rate
to more than 100 times the experimentally obtained value.


\textbf{Conclusions:} This paper makes the first application of the
\emph{classical field method} to a realistic three dimensional problem, and 
produces results which could not have been generated by any other method in 
current use.  Our detailed numerical results require only \emph{a single run} 
of the simulation, avoiding time-consuming ensemble averages. Previous
calculations \cite{Band2000a,Vogels2002a,Bach2002a,Koehler2002a} have been
restricted to regimes in which Bosonic stimulation is not important, and have
only been able to determine more limited information, whereas our method can
handle very large ($ \approx 40 $\%) outscattered fractions.  

The only significant restriction on our method is the computer memory required
to represent the system as it expands in space. For the situation considered
here this is not a problem. Collisions with smaller out-scattered fractions
would be more difficult to simulate because of the high ratio of virtual to
real particle occupation, but the method would still be valid if sufficiently
large ensembles of initial conditions were used, whereas for large
out-scattered fractions, a single realization is sufficient for most measurable
quantities.

New qualitative features found are: i) The existence of \emph{quantum
turbulence}, which should be easily detectable experimentally, using for
example measurement of the density along a slice with the same techniques used
to observe similar slices through clouds containing vortex lattices
\cite{Raman2001a,Haljan2001a};\quad ii) The suppression of the modulational
instability which a simple Gross-Pitaevskii picture would predict;\quad iii)
The reduction in correlations expected between scattered atoms of opposite 
momentum to a rather small value.
\acknowledgments
The authors would like to thank A. S. Bradley for helpful discussions. This work
was supported by the New Zealand Marsden Fund under contract number PVT-202.
\vspace{-0.6cm}
\bibliographystyle{apsrev}
\bibliography{HaloAugust}

\end{document}